\documentclass[aps,showpacs,prb,twocolumn]{revtex4}
\usepackage{graphicx}

\begin{document}

\title{Magnetism and superconductivity in strongly correlated CeRhIn$_5$}

\author{Tuson Park$^{1,2}$ and J. D. Thompson$^2$}
\affiliation{$^1$ Department of Physics, Sungkyunkwan University, Suwon 440-746, Korea \\ $^2$Los Alamos National Laboratory, Los Alamos, New Mexico 87545, USA}
\begin{abstract}
Specific heat studies of CeRhIn$_5$ as functions of pressure and magnetic field have been used to explore the relationship between magnetism and unconventional superconductivity, both of which involve the 4f electron of Ce. Results of these studies cannot be understood as a simple competition for Fermi-surface states and require a new conceptual framework.
\end{abstract}

\maketitle

\section{Introduction}
When Ce is added as a dilute magnetic impurity into a non-magnetic host, antiferromagnetic exchange $J$ between band electrons of the host and the localized 4f electron of Ce$^{3+}$ produces a magnetic-singlet ground state in the zero-temperature limit. This many-body process, the Kondo effect, renormalizes the effective mass of itinerant charge carries that is reflected in a large specific heat Sommerfeld coefficient per mole Ce  and that scales as     $\gamma \propto 1/T_K$~\cite{bader}, where the Kondo temperature $T_K \approx  N(E_F)exp[-1/2JN(E_F)]$ and $N(E_F)$ is the density of electronic states of the host~\cite{hewson}. The resulting Fermi volume of the interacting system is 'large', i.e., counts both band electrons of the host and the 4f electron of Ce.  In contrast to the well-understood impurity limit, a periodic array of Kondo atoms is much more complex. In this dense limit, the Kondo effect as well as interactions, such as the indirect Ruderman-Kittel-Kasuya-Yosida (RKKY) interaction, between local moments are present. With both Kondo and RKKY interactions depending on $J$, their interplay can be tuned by chemical environment, magnetic field and pressure to give magnetically ordered, superconducting or paramagnetic ground states of the Kondo lattice~\cite{yang08}. If Kondo screening dominates, then the Fermi volume of the heavy quasiparticles also will be large~\cite{hvLreview}. The surface of this large Fermi volume is susceptible to instabilities, such as superconductivity and/or formation of a spin-density wave. CeCu$_2$Si$_2$, the first heavy fermion superconductor~\cite{steglich79}, is an example. Its large Fermi volume of itinerant heavy quasiparticles forms unconventional superconductivity, a spin-density wave or both, depending on chemical stoichiometry~\cite{grewe91}. In the sense that these broken symmetries are the consequence of a Fermi-surface instability, the same (itinerant) electrons are involved in both superconductivity and magnetism. On the other hand, if the interaction between local moments dominates, local-moment magnetic order develops in a 'small' Fermi volume that does not count the localized electrons, and magnetic entropy in the ordered state will be close to that expected from the degeneracy of the localized electrons. CeCu$_2$Ge$_2$ is an example of the localized limit; localized 4f electrons order antiferromagnetically near 4K in a well-defined crystal-field doublet manifold. CeCu$_2$Ge$_2$ becomes superconducting at very high pressures~\cite{jaccard95}, where the Kondo effect dominates, and, consequently, magnetism and superconductivity arise from electrons with very different characters, one localized and the other itinerant.

The dense Kondo system UPd$_2$Al$_3$ has been proposed to exhibit both localized and itinerant characters of its three 5f electrons, with two of the 5f electrons being localized and ordering antiferromagnetically and one of the 5f electrons being itinerant and participating in coexisting unconventional superconductivity~\cite{caspary93}. From a theoretical perspective, division of electronic orbitals into localized and itinerant components is possible depending on the competition between intra-atomic Coulomb interactions and anisotropic hybridization of 5f electrons~\cite{efremov04}. Though this mechanism may account for the behavior of U's 5f electrons in UPd$_2$Al$_3$, it remains to be shown that such competition could lead to simultaneously localized and itinerant characters of a single f electron. In addition to the relationship between magnetism and superconductivity, the issue of itinerant or localized character of electrons is important for interpreting the nature of quantum criticality that emerges as the Neel transition of a Kondo lattice is tuned to zero temperature. The Hertz-Millis-Moriya theory is constructed to account for quantum-critical behavior arising from a zero-temperature spin instability of a large Fermi volume~\cite{hvLreview}, and, indeed, transport and thermodynamic properties of CeCu$_2$Si$_2$ that emerge as its spin-density transition is tuned to $T=0$ can be interpreted in this framework~\cite{steglich96}. On the other hand, behaviors of the Kondo lattice systems YbRh$_2$Si$_2$~\cite{paschen04} and CeCu$_{6-x}$Au$_x$~\cite{schroder98} near their quantum-critical points have led to qualitatively different models that assume localized 4f electrons~\cite{hvLreview}, and in contrast to CeCu$_2$Si$_2$, neither of these systems exhibits superconductivity near their quantum-phase transition~\cite{gegenwart08}. The distinction, however, between localized and itinerant magnetic phases is not always straightforward, with some theories arguing that they are distinct phases and others for a continuous transition from localized to itinerant antiferromagnetism, i.e., from small to strong Kondo coupling, as a function of some tuning parameter that also accesses the quantum-critical regime.

CeRhIn$_5$ is an example of complexity posed by a Kondo lattice. At atmospheric pressure, localized 4f electrons in CeRhIn$_5$ order antiferromagnetically at 3.8~K. Studies of dilute Ce in LaRhIn$_5$ show that the Kondo-impurity temperature $T_K\approx 0.15K$ is more than one order of magnitude smaller than T$_N$ of CeRhIn$_5$ at atmospheric pressure~\cite{yang08}, consistent with local moment order and the small Fermi surface deduced from de Haas-van-Alphen (dHvA) experiments~\cite{shishido05}. Whereas, the small impurity $T_K$ would imply a huge Sommerfeld coefficient of $>35$J/mol$\cdot$K$^2$ and negligible Kondo compensation of the local moment, the experimental   for CeRhIn$_5$ above $T_N$ is about 0.4~J/mole$\cdot$K$^2$; magnetic entropy below $T_N$ is $\approx 0.3$Rln2; and the order moment of $0.79 \mu _B$ is reduced by about $10\%$ from that of Ce in a crystal-field doublet without the Kondo effect~\cite{hegger00, anna04}. Together, these observations are inconsistent with naive expectations and suggest that electrons in the Kondo lattice are neither simply localized nor itinerant. This is reflected as well in the response of CeRhIn$_5$ to pressure. Modest pressure induces a phase of coexisting superconductivity and local-moment magnetic order that evolves at higher pressures to a superconducting state without magnetic order~\cite{mito03,kawasaki03,tuson06}. Applying a magnetic field, however, to this higher pressure superconducting state induces magnetic order that coexists with superconductivity to a critical pressure where the field-induced magnetic order is tuned to $T=0$~\cite{tuson06,knebel06}. At this critical pressure, transport and thermodynamic properties exhibit strong deviations from those of a heavy Landau Fermi liquid~\cite{tuson08}, and de Haas-van Alphen frequencies of principal orbits increase sharply~\cite{shishido05}, as if signaling a transition from small-to-large Fermi volume, again emphasizing ambiguity in the nature of electrons responsible for superconductivity, magnetism and non-Fermi-liquid responses. As discussed below, specific heat studies of CeRhIn$_5$ as a function of temperature, pressure and magnetic field explicitly reveal this ambiguity and allow exploration of its consequences on the symmetry of the superconducting gap.

\section{Experimental}
Single crystals of CeRhIn$_5$ were grown from In flux as described elsewhere. Powder x-ray diffraction confirmed the HoCoGa$_5$ tetragonal structure, which can be viewed as layers of CeIn$_3$ and RhIn$_2$ stacked sequentially along the c-axis. Electrical resistance measurements indicated the high quality of these crystals, with a resistivity ratio $\rho$(300K)/ $\rho$(0K) $>$ 400 and a residual resistivity of $\approx 40$ n$\Omega \cdot$cm.  Specific heat measurements were performed in a hybrid clamp-type pressure cell made of Be-Cu/NiCrAl in which there was a Teflon cup that contained a silicone-fluid pressure transmitting medium, a small piece of Sn, whose inductively measured superconducting transition served to determine pressure in the cell at low temperatures, and a sample of CeRhIn$_5$. The specific heat was determined by ac calorimetry. In this technique, an alternating current through a heater, attached to one side of the sample, induced an oscillating change in temperature $T_{ac}$, detected on the other side of the sample by a calibrated Au-Fe thermocouple and that is inversely proportional to the sample's specific heat $C \propto 1/T_{ac}$. This technique is not absolute because the amplitude and phase of temperature oscillations are influenced by heat transported to the surrounding pressure medium. Approximately absolute values of the specific heat of CeRhIn$_5$ were obtained by scaling the measured specific heat to that determined earlier by an adiabatic technique~\cite{fisher02}. The pressure cell was attached to $^3$He cryostat that fit into a 9-T superconducting solenoid.

\section{Results}
\subsection{Phase Diagram}

Figure 1 displays representative specific heat measurements of CeRhIn$_5$ under pressure and zero magnetic field. Specific heat data at 1.15~GPa, which are denoted by down triangles, show a magnetic transition to an incommensurate antiferromagnetic (ICM) phase with Q=(0.5, 0.5, 0.297) at 3.62~K and a pressure-induced superconducting transition at a much lower temperature ($T_c =0.36$~K). With further increasing pressure, the magnetic transition temperature $T_N$ decreases, while $T_c$ progressively increases. At 2.05~GPa (circles), there is a large specific heat discontinuity at 2.24~K due to a superconducting phase transition, but evidence for long-range magnetic order is completely absent. Even though $T_N$ sensitively depends on pressure, the shape and full width at half-maximum (FWHM) of the magnetic transition is almost independent of pressure, indicating that the nature of the magnetic transition has not been changed under these hydrostatic pressure environments. This conclusion is consistent with neutron-scattering measurements under pressure~\cite{anna04}.

\begin{figure}[tbp]
\centering  \includegraphics[width=7.5cm,clip]{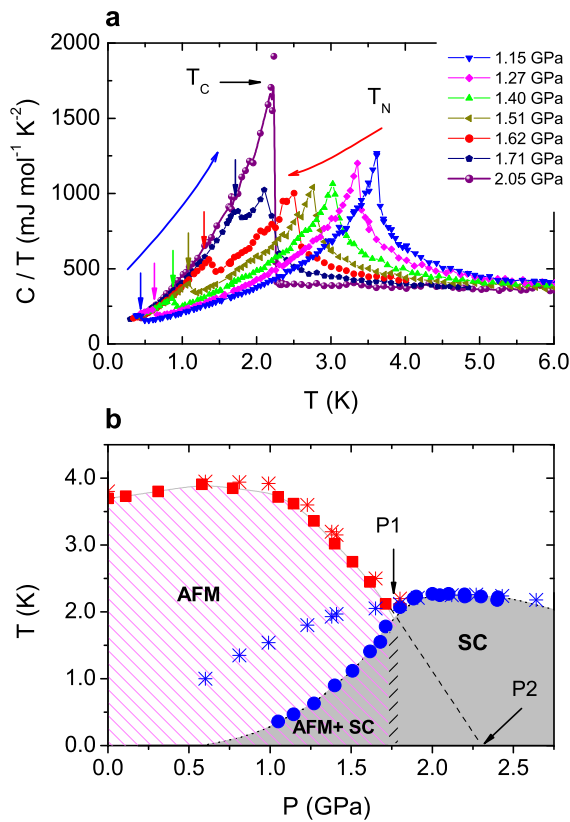}
\caption{(color online)  a) Specific heat of CeRhIn$_5$ as a function of temperature at zero magnetic field. $T_c$, which is assigned as the mid point of the jump in specific heat due to superconductivity, is marked by arrows. b) Temperature-pressure phase diagram at zero field. $T_c$ and $T_N$ are determined by specific heat measurements (filled symbols) and by electrical resistivity measurements (crosses).}
\label{figure1}
\end{figure}

Under applied pressure, CeRhIn$_5$ has been shown to possess two critical points (see Fig.~1b)~\cite{tuson06}. The first occurs at 1.75~GPa~(=P1), where the magnetic and superconducting transition temperatures become equal in the absence of an applied magnetic field and the second at P2 where a field-induced magnetic transition is tuned to $T=0$~\cite{tuson06,knebel06}. The magnetically ordered phase is completely suppressed for $P>P1$; whereas, magnetic and superconducting phases coexist on a microscopic scale for $P<P1$~\cite{mito03, kawasaki03}. The abrupt suppression of AFM at P1 has been interpreted as an indication of a weakly first-order phase transition between a state in which magnetic order and superconductivity coexist to a purely superconducting phase ~\cite{flouquet05}. When extrapolated below 1.15~GPa, the bulk $T_c$ determined from $C_p$ is zero near 0.5~GPa at which $T_N$ starts to decrease, indicating an interplay between magnetic and superconducting phases. This conclusion is consistent with our specific heat measurements that do not show any hint of bulk superconductivity above 70 mK at ambient pressure, but is different from Ref.~\cite{jp07} that claimed bulk superconductivity near 110~mK.

In Fig.~1b, we also plot $T_c$ and $T_N$ determined from electrical resistivity measurements. At ambient pressure, there is a slight decrease in resistivity below 50~mK possibly due to incomplete filamentary superconductivity (not shown). Unlike the bulk $T_c$ that increases gradually with pressure, the $T_c$ determined from zero resistance sharply increases above 0.5~GPa, resulting in a large football-shaped difference curve in the two $T_c$s. For pressures above P1, the two $T_c$s become almost equal. A difference between the bulk and transport $T_c$ also has been reported in a sister compound CeIrIn$_5$~\cite{petrovic01}. Polishing, etching, and heavy-ion irradiation of the surface of CeIrIn$_5$ do not affect the two $T_c$s, ruling out the possibility of surface superconductivity. Chemical substitution, however, narrows the difference between the two $T_c$s, suggesting that a strain may exist locally to affect superconductivity of CeIrIn$_5$ and chemical disorder relieves the local strain to make the system more homogeneous~\cite{bianchi02}. In CeRhIn$_5$, the difference between bulk and resistive $T_c$s is reversible with decreasing pressure, which indicates an intrinsic response to the coexistence of magnetic and superconducting orders and not an effect of internal strain. A strain scenario would require large pressure inhomogeneity, at least of order 0.5~GPa even at ambient pressure. Such a large pressure gradient is at odds with independent studies of the hydrostaticity of our pressure medium in which there is a maximum gradient of 0.01 GPa even when the silicone fluid freezes into a glassy state~\cite{sidorov}. In addition, the coincidence of bulk and resistive transitions for $P>P1$ also is inconsistent with a strain scenario. An alternative interpretation is that the resistive transition arises from magneto-elastic coupling that initially induces filamentary superconductivity at antiferromagnetic domain walls before the sample bulk superconducts at lower temperature.

\subsection{Coexistence of magnetism and superconductivity below P1}

Though specific heat measurements establish two phase transitions, bulk superconductivity and magnetic order, at pressures below P1, and NMR measurements show that these orders coexist on the scale of a unit cell, neither is able to determine the nature of the electronic state out of which these orders develop. Neutron diffraction at pressures to 1.6~GPa show that the size of the ordered moment ($0.79~\mu _B$) is only weakly pressure dependent and close to that of Ce$^{3+}$ in a crystal field doublet, indicating that magnetism in CeRhIn$_5$ arises from localized Ce 4f spins~\cite{anna04}. The localized nature of Ce's 4f electron in the AFM state is consistent with de Haas-van Alphen (dHvA) experiments on Ce-doped LaRhIn$_5$, where quantum-oscillation frequencies reflecting a small Fermi surface are independent of Ce-doping concentration up to stoichiometric CeRhIn$_5$~\cite{harrison04}. These results argue for local-moment order mediated by the RKKY interaction. As a function of pressure, the dHvA frequencies of CeRhIn$_5$ are unchanged for $P<P1$~\cite{shishido05}, i.e., the f-electron remains localized so that the coexistence of magnetic order and superconductivity should not be viewed as a simple competition for electronic states at the Fermi energy. On the other hand, inspection of specific heat data plotted in Fig.~1a shows that superconductivity develops in the antiferromagnetic state in which $C/T$ just above $T_c$ is a substantial fraction of $C/T$ just above $T_N$, that is, Neel order has not dramatically suppressed the effective mass of quasiparticles that form Cooper pairs. Such enhanced effective mass must come from hybridization of f- and band electrons since there is no enhancement of the effective mass of quasiparticles in non-magnetic LaRhIn$_5$~\cite{harrison04}. These observations lead to a dichotomy in which the 4f electron produces both local moment magnetic order and participates in creating heavy mass quasiparticles that form superconductivity.  We have used specific heat measurements as a function of pressure and magnetic field to explore the relationship between magnetism and superconductivity in the coexistence region below P1.

\begin{figure}[tbp]
\centering  \includegraphics[width=7.5cm,clip]{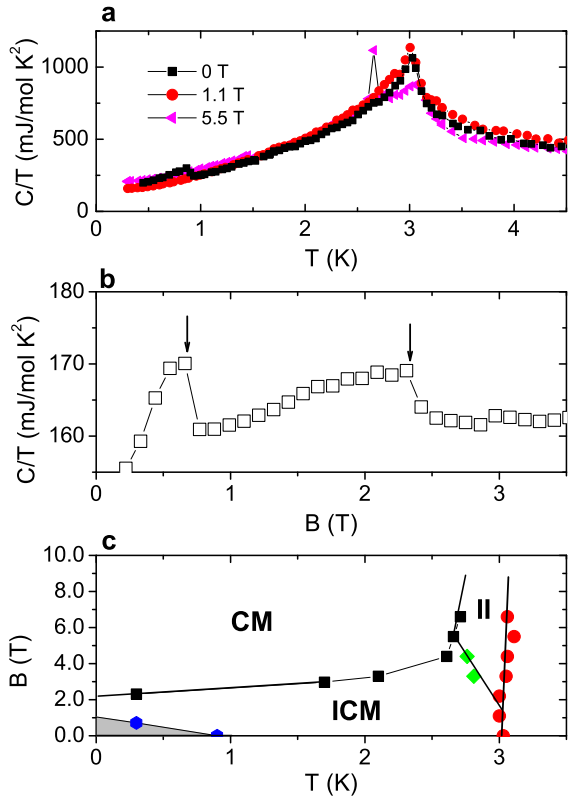}
\caption{(color online)   Specific heat for $B \perp$ c-axis at 1.4~GPa. a) Specific heat divided by temperature $(C/T)$ is plotted as a function of temperature for 0, 1.1, and 5.5~T. b) Magnetic field dependence of $C/T$ at 0.3~K. Arrows indicate superconducting and magnetic transition at 0.66 and 2.3~T, respectively.  Data were taken with increasing field. c) $B-T$ phase diagram. Tuning magnetic field and temperature reveals three distinct magnetic phases of  incommensurate (ICM) with \textbf{Q}=(0.5, 0.5, 0.297), commensurate (CM) with \textbf{Q}=(0.5, 0.5, 0.25), and another incommensurate (II) magnetic order~\cite{raymond}. Solid lines are guides to eyes.}
\label{figure2}
\end{figure}

Figure 2a displays specific heat data for CeRhIn$_5$ at 1.4~GPa and in magnetic fields of 0, 1.1 and 5.5~T. In zero field, a magnetic transition occurs at 3.0~K and then a superconducting transition follows at 0.9~K. Superconductivity is rapidly suppressed by an applied magnetic field, while the magnetic order is robust. At 5.5~T, there are two closely spaced specific heat anomalies. The sharpness of the second peak at lower temperature is indicative of its first order nature. Magnetic field dependence of $C/T$ at 0.3~K is plotted in Fig.~2b. There are two anomalies at 0.6 and 2.3~T that correspond to superconducting and field-induced magnetic transition, respectively. From data such as these, we construct the field-temperature phase diagram plotted in Fig.~2c that summarizes the evolution of electronic phase of CeRhIn$_5$ at 1.4~GPa. The $B-T$ phase diagram at this pressure is essentially identical to that at 1~bar~\cite{cornelius01}, except for the presence of superconductivity at low-temperatures and low-fields (shaded area). Neutron-diffraction measurements at 1~bar show that the zero-field transition is to an incommensurate (ICM) structure (0.5, 0.5, 0.297)~\cite{bao}; whereas, the low-temperature, field-induced transition is commensurate (CM) at (0.5, 0.5, 0.25)~\cite{raymond}. Even though the zero-field magnetic transition temperature is suppressed from 3.8~K at 1~bar to 3.0~K at 1.4~GPa, the critical field $(B^*)$ required for the field-induced spin-reorientation transition from ICM to CM is almost independent of pressure up to this pressure: $B^*= 2.0$~T at 1~bar and 2.2~T at 1.4~GPa and 0~K.

\begin{figure}[tbp]
\centering  \includegraphics[width=7.5cm,clip]{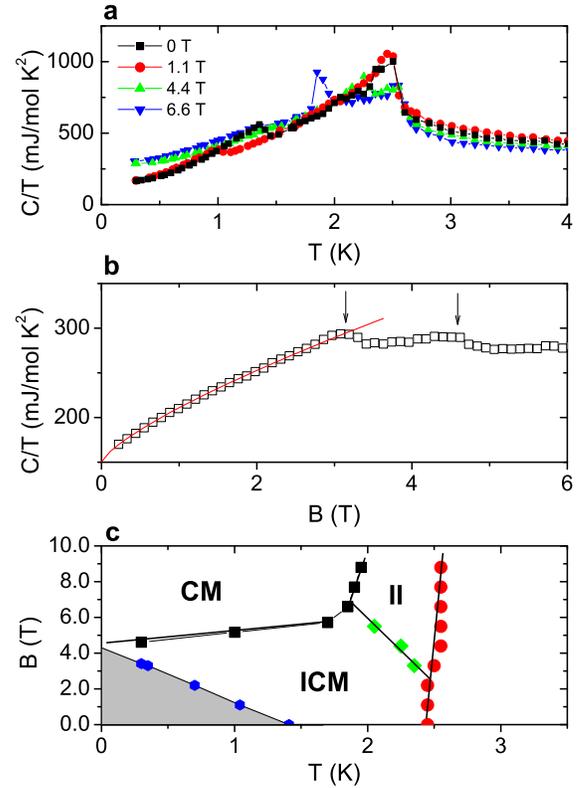}
\caption{(color online)   Specific heat for $B \perp$ c-axis at 1.62~GPa. a) Specific heat divided by temperature $(C/T)$ is plotted as a function of temperature for 0, 1.1, 4.4, and 6.6~T. b) Magnetic field dependence of $C/T$ at 0.3~K. Arrows indicate superconducting and magnetic transition at 0.66 and 2.3~T, respectively. The solid line is a least-square fit to $C/T =\gamma+\beta B^n$ where best result was obtained with $n=0.75$. c) $B-T$ phase diagram based on the specific heat measurements.}
\label{figure3}
\end{figure}

Figure 3a shows specific heat at 1.62~GPa, which is close to the
critical boundary P1 and where $T_N$ is depressed to 2.5~K and
$T_c$ is increased to 1.4~K. Like results at 1.4~GPa, the
specific heat anomaly at $T_N$ splits into two peaks under
magnetic field. However, the critical field $B^*$ that is
required for the spin reorientation transition from ICM to CM has
jumped to 4.4~T, a factor of 2 larger than that at 1.4~GPa. At
1.71~GPa, even closer to P1, $B^*$ is 6.7~T~\cite{tuson08a}. The
field $B^*$ is almost constant ($\approx 2$~T) up to 1.4~GPa,
where the superconducting upper critical field $B_{c2}$ is much
less than $B^*$. For $P\geq 1.5$~GPa, where $B_{c2}$ sharply
increases above 2~T, $B^*$ increases in step with the increase in
$B_{c2}$ and does not cross $B_{c2}$, indicating that
superconductivity coexists with the ICM phase but competes
against the CM phase. Though $B^*$ and $T_N$ change
quantitatively with pressure, the relationship among magnetic
phases remains unchanged from $P=0$ to $P<P1$, reflecting little
change in the localized nature of the 4f electron over this
pressure range; nevertheless, the superconducting transition
temperature increases by nearly an order of magnitude. We note
that CeRhIn$_5$ is not alone in exhibiting this apparent
localized/itinerant dichotomy. Its isostructural relative
CeCoIn$_5$ is a heavy-fermion superconductor with a 'large' Fermi
surface~\cite{shishido02}. Replacing approximately 1~at/\% In
with Cd induces a phase of coexisting antiferromagnetism and
unconventional superconductivity~\cite{pham, urbano}, and in
several respects the temperature - Cd doping phase diagram is a
mirror image $T-P$ phase diagram of CeRhIn$_5$.
Neutron-diffraction measurements on 1~at/\% Cd in CeCoIn$_5$ find
magnetic scattering intensity developing below $T_N$ that is
arrested with the onset of superconductivity~\cite{nicklas07},
clearly demonstrating a coupling between the two orders. It would
be interesting to see if this same coupling were apparent in
CeRhIn$_5$ at pressures less than P1 and if the Fermi surface of
Cd-doped CeCoIn$_5$ were 'small'.

Magnetic field dependence of the density of sates (DOS) can be used to explore the nature of the superconducting (SC) order parameter. In conventional superconductors, where the SC gap is finite on the whole Fermi surface, the density of states is proportional to magnetic field because the DOS comes from states inside the normal core of vortices and the density of magnetic vortices is proportional to $B$. In unconventional superconductors with nodes, where the SC gap becomes zero on parts of the Fermi surface, the DOS mainly comes from extended quasiparticle states along the nodal directions and is expected to show a $B^{1/2}$ dependence~\cite{sigrist}. Figure~3b shows $C/T$ as a function of magnetic field at 1.62~GPa. There are two anomalies that are associated with SC transition and spin-reorientation transitions at 3.4 and 4.6~T, respectively. The solid line is a least-squares fit to a power law, $C/T =\gamma +\beta B^n$, in the SC state, where $n=0.75$ produces the best fit. The obtained exponent at this pressure does not conform to either a d-wave or an s-wave scenario. It is instructive to note that the SC upper critical field at this pressure, as shown in Fig.~3c, also deviates from the conventional Werthamer-Helfand-Hohenberg (WHH) model of orbital pair breaking~\cite{whh66}, showing a linear temperature dependence down to the lowest experimental temperature. The nonconformity of the field-dependent specific heat and the temperature-dependent $B_{c2}$ may stem from the presence of incommensurate magnetic ordering, which could allow a superconducting order parameter other than spin-singlet Cooper pairing.

\subsection{Field-induced magnetism and superconductivity}

\begin{figure}[tbp]
\centering  \includegraphics[width=7.5cm,clip]{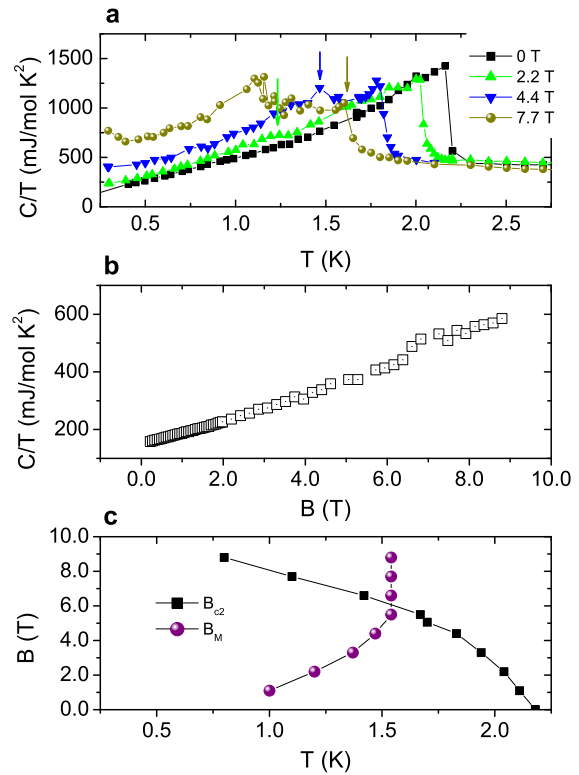}
\caption{(color online)  a) Specific heat measurements at 1.88~GPa and for $B \perp$ c-axis. a) $C/T$ is plotted as a function of temperature for 0, 2.2, 4.4, and 7.7~T. Arrows indicate $B-$induced magnetic transition temperature. b) Magnetic field dependence of $C/T$ at 0.3~K. c) Field-temperature phase diagram of CeRhIn$_5$ at 1.88~GPa.}
\label{figure4}
\end{figure}
When pressure is higher than 1.75~GPa, evidence for a magnetic transition temperature $T_N$ is abruptly lost (see Fig.~1b), suggesting a first-order phase transition at P1 from a coexisting phase of superconductivity and magnetism to a solely superconducting phase. The lack of magnetism above P1 might be understood if conduction electrons required for the indirect RKKY interaction were completely gapped by the development of superconductivity; however, the superconducting gap is non-zero on parts of the Fermi surface at these pressures, as evidenced by a $T^3$ dependence of the nuclear relaxation rate~\cite{kawasaki03}. Figure~4a displays specific heat measurements at 1.88~GPa that represent the high-pressure behavior of CeRhIn$_5$ for $P1 < P < P2$ (Fig.~1b). At zero field, a jump in the specific heat occurs at 2.2~K due to the onset of bulk superconductivity, but there is no detectable feature that would indicate a magnetic transition at a lower temperature, e.g, near 1~K where a simple extrapolation of the Neel boundary would suggest that one might appear. When subjected to an external magnetic field as low as 1.1~T (not shown), however, a second peak in $C/T$ appears in the superconducting state. This low-temperature anomaly is enhanced with increasing magnetic field: the field-induced transition temperature $T_M$ increases and the peak anomaly becomes more conspicuous with increasing field, as expected if the anomaly signaled the development of long range magnetic order. It could be speculated that the field-induced transition in the superconducting state is similar to the spin-reorientation transition from ICM to CM state for $P < 1.75$~GPa. However, the fact that the spin-reorientation transition $B^*$ is larger than 6~T near 1.7~GPa but a field $B_M$ as low as 1.1~T is sufficient to induce magnetism at 1.88~GPa indicates that the two critical fields $B^*$ and $B_M$ are of different natures.

Figure 4b shows $C/T$ as a function of magnetic field at 0.3~K and 1.88~GPa. The lack of feature at the field-induced transition $B_M$ (1~T) is due to an immeasurably small change in specific heat at the transition in these field-swept data.  As seen in these data, the field dependence of $C/T$ is very different from that at 1.62~GPa, now showing a linear$-B$ variation. Even though the linear dependence is consistent with an s-wave order parameter, it is premature to conclude that a dramatic change in the nature of superconductivity has occurred upon crossing the critical pressure P1 (=1.75~GPa) because, as noted above, a $T^3$ dependence of $1/T_1$ in zero field implies a d-wave order parameter for $P>P1$. Instead, the linear$-B$ dependence may be influenced by the presence of field-induced magnetic order in the Abrikosov state, but to our knowledge there is no theoretical prediction for what might be expected in this case.

The relation between superconductivity and field-induced magnetic order, deduced from data such as shown in Fig.~4a, is plotted in Fig.~4c and is characteristic of this relationship for pressures greater than P1 but less than P2. As seen here, the phase boundary $B_M(T)$ extends into the normal state above $B_{c2}$. It is interesting that the upper critical field boundary becomes linear in field just below the temperature where $B_M(T)$ crosses $B_{c2}(T)$ and that $B_M(T)$ has qualitatively difference temperature dependences in and out of the superconducting state, both suggesting a coupling of the two order parameters. The near field-independence of $B_M(T)$ above $B_{c2}(T)$ is reminiscent of the response of $T_N$ to field for $P<P1$. This similarity further suggests that the field-induced magnetic order above P1 is of the local-moment type, probably incommensurate, that is found below P1. Neutron-diffraction studies would be very useful to confirm or deny this speculation. We note that the relation between field-induced magnetic order and superconductivity found here is distinctly different from what is observed in CeCoIn$_5$ where field-induced spin-density order exists only inside its Abrikosov phase~\cite{kenzelmann}

\begin{figure}[tbp]
\centering  \includegraphics[width=7.5cm,clip]{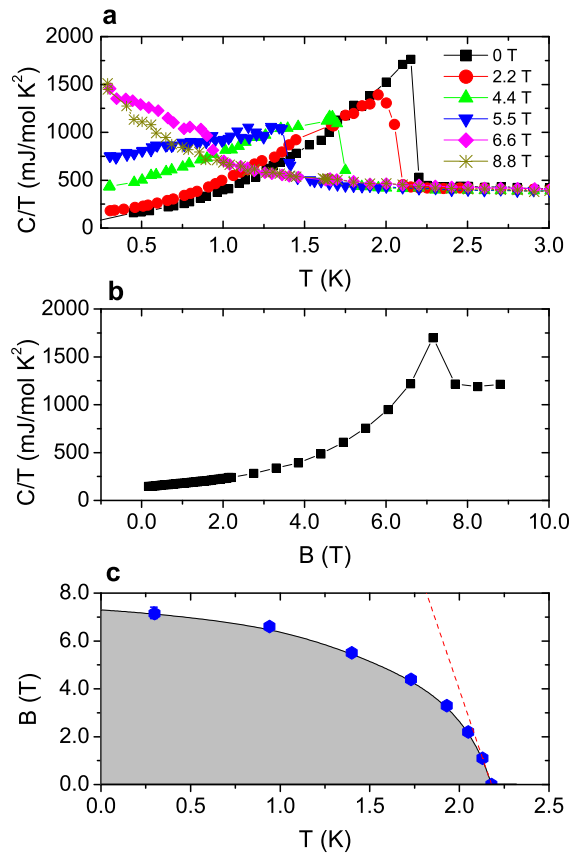}
\caption{(color online) Specific heat measurements at 2.45~GPa and for $B \perp$ c-axis. a) $C/T$ as a function of temperature for several fields. b) $C/T$ at 0.3~K plotted as a function of magnetic field. c) Field-temperature phase diagram. The dashed line is a linear fit to $B_{c2}$ near $T_c$ and gives a slope of  22~T/K.}
\label{figure5}
\end{figure}

At pressures higher than 2.35~GPa (=P2), the field-tuned quantum-critical point~\cite{tuson06}, evidence for field-induced magnetic order disappears. This is illustrated in Fig.~5a where we plot specific heat data as a function of temperature for CeRhIn$_5$ at 2.45~GPa, which is slightly above P2. With increasing field, $T_c$ is initially suppressed at a rate of 45.5~mK/T and at the highest field, $C/T$ continues to increase to the lowest temperatures without the appearance of a field-induced anomaly. If the upper critical field were determined by orbital pair-breaking effects, then the zero-temperature upper critical field,  estimated from $B_{c2}(0) = -0.73T_c (dB_{c2}/dT) = 35$~T, far exceeds the experimental value of 7.3~T (see Fig.~5c). In addition to orbital pair breaking, in a spin-singlet superconductor, the upper critical field can be limited by field-induced alignment of spins, namely Pauli paramagnetic pair breaking~\cite{clogston}. The fact that the observed $B_{c2}$ is smaller by a factor of 5 than the estimated orbital critical field indicates that Pauli pairing breaking dominates, which contrasts with results below P1 (for example, Fig.~3c) where the extrapolated $B_{c2}(0)$ is larger than that determined by orbital effects. As shown in Fig.~5b, the specific heat anomaly at $B_{c2}$(0.3~K) is nearly symmetric, typical of a first order phase transition, and in contrast to the second-order like anomaly at $T_c(B)$ in lower fields (Fig.~5a).  Such a first order transition is expected for a Pauli-limited $B_{c2}$~\cite{maki}. In this limit, we might expect the emergence of an inhomogeneous superconducting phase predicted by Fulde and Ferrell~\cite{fulde} as well as by Larkin and Ovchinnikov~\cite{larkin}. Though evidence for such a phase has been found in CeCoIn$_5$~\cite{bianchi03, radovan}, which has a similar ratio of Pauli to orbital critical fields, we have not yet resolved a feature in the high-field, low-temperature specific heat of CeRhIn$_5$ that might be associated with this inhomogeneous superconducting phase.

\section{Discussion}

In conventional magnetic superconductors where magnetism and superconductivity coexist, localized f electrons from a rare-earth element contribute to a local magnetic order but decoupled itinerant electrons are responsible for conventional superconductivity. In strongly correlated systems, in contrast, the same electrons can be responsible for both magnetism and unconventional superconductivity: Cu 3d electrons in high-$T_c$ cuprates or 4f (5f) electrons from rare-earth (or actinide) elements in heavy-fermion superconductors. Recently discovered Fe-pnictide superconductors also may belong to the class of electronically correlated superconductors~\cite{haule08}, where Fe 3d electrons are responsible for both superconductivity and magnetic order~\cite{christiansong08}.

Understanding the coexisting phase of CeRhIn$_5$ is complex even though the low-pressure AFM state and high-pressure SC state can be interpreted as arising from dominantly localized and itinerant characters of Ce's 4f electron, respectively. The abrupt increase in dHvA frequencies near 2.35~GPa of a dominant quasi-two dimensional sheet~\cite{shishido05} suggests a localized to delocalized transition in the 4f character, perhaps signaling a selective Mott transition at this quantum critical point~\cite{pepin}. Because all Fermi-surface sheets are not accounted for in these studies, it is not possible to conclude that the 4f electron has become completely delocalized to form a large Fermi surface at high pressures; however, these higher frequencies do correspond closely to those of isostructural, heavy-fermion superconductor CeCoIn$_5$ whose Fermi-surface topology and size agree well with band calculations that assume itinerancy of the 4f electrons~\cite{haga01}.

\begin{figure}[tbp]
\centering  \includegraphics[width=7.5cm,clip]{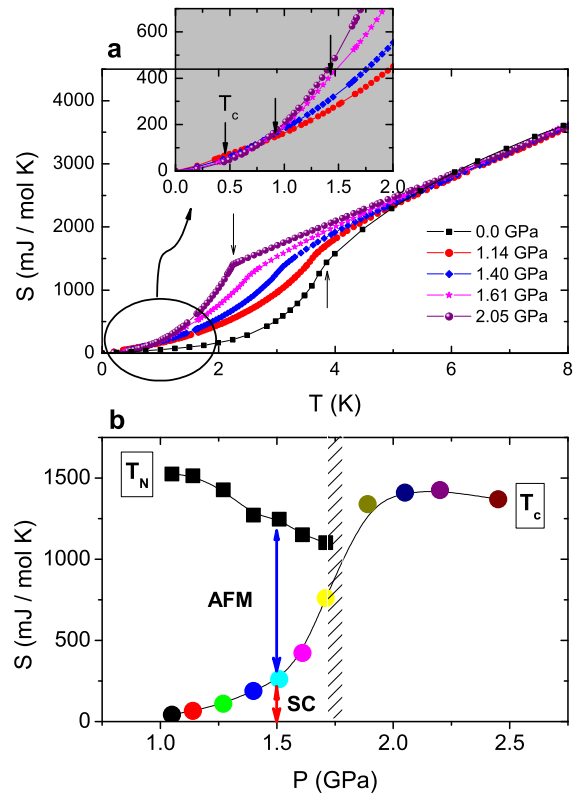}
\caption{(color online)  a) Entropy as a function temperature for several pressures at zero magnetic field. Inset: Magnification of low$-T$ superconducting transition. b) Comparison of entropy values at $T_N$ (solid squares) and $T_c$ (solid circles) as a function of pressure. Hashed area marks a zero-field critical pressure P1 (=1.75~GPa).}
\label{figure6}
\end{figure}

In the absence of a direct probe of the 4f electron, inferences of its contributions to magnetism and superconductivity are provided by the evolution of magnetic entropy with pressure -- see Fig.~6. At ambient pressure, there is a sharp kink at $T_N$ (marked by an arrow) and about $70\%$ of the Rln2 spin entropy is recovered near 8~K, where neutron measurements show the onset of short range spin-spin correlations~\cite{bao02}. Even though the ground state of CeRhIn$_5$ is sensitive to pressure, the entropy below 8~K is almost independent of the applied pressure and ground state. Figure~6b shows how entropy at $T_c$ and $T_N$ evolves as a function of pressure. At 1.05~GPa, entropy of the Neel order dominates, and entropy associated with superconductivity is negligible. With increasing pressure, entropy due to magnetic order decreases and the 'lost' entropy is transferred to SC entropy. When pressure is higher than 1.75~GPa, entropy is completely in the superconducting channel. The transfer of spin entropy from magnetism to superconductivity provides strong evidence that the single Ce 4f electron does participate in the two disparate broken symmetries and, further, that superconductivity has a magnetic origin.

The superconducting order parameter in the high-pressure phase of CeRhIn$_5$ is most likely d-wave. NMR measurements show a $T^3$ dependence below $T_c$ in the spin-lattice relaxation rate\cite{kawasaki03}; there is no Hebel-Slichter peak below $T_c$; the upper critical field is Pauli limited; and, as shown in Fig.~7, field-rotation specific heat measurements under pressure find a four-fold modulation when magnetic field is rotated within Ce-In plane. We note that these experimental evidences also characterize CeCoIn$_5$, where a d-wave order parameter is established~\cite{roman01,izawa,greene08}.

\begin{figure}[tbp]
\centering  \includegraphics[width=7.5cm,clip]{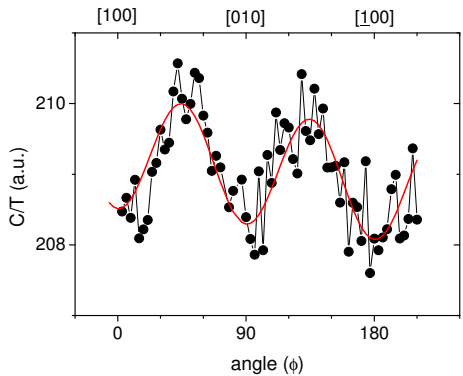}
\caption{(color online)  In-plane, field-angle dependent ac specific heat
divided by temperature for CeRhIn$_5$ at 2.3 GPa. Measurements
are made a fixed temperature of 325~mK and fixed magnetic field
of 0.9~T, which is well below the upper critical field. The solid
line is a least-squares fit to a four-fold modulation model:
$C(\phi) = C(0) + A (1-A4cos( 4\phi )) + B\phi$ , where C(0)=200,
A=9.3, A4=0.0856, B=-0.00243. We tentatively attribute the
linear-in angle, $B\phi$, contribution to slight misalignment of
the crystal.} \label{figure7}
\end{figure}

In the coexisting phase, the nature of the superconducting order parameter is less clear. The presence of magnetism as well as proximity to a magnetic critical point has prompted various theoretical predictions, ranging from a generalized d-wave with eight nodal points to a gapless p-wave superconductivity~\cite{bang04, fuseya03}. Experimentally, NQR measurements show that $1/T_1$ crosses over from a $T^3$ to $T-$linear dependence deep in the SC state at 1.6~GPa, possibly indicating a gapless p-wave order parameter. Specific heat measurements, however, observe a finite $T=0$ density of states in the coexisting phase, which can also lead to a $T-$linear crossover. In addition, field-rotation specific heat measurements have revealed a four-fold oscillation, which is consistent with a d-wave order parameter in the coexisting phase~\cite{tusonPRL}. The dependence on magnetic field of $C/T$ in the coexisting phase, however, cannot be explained by a d-wave alone, requiring an interplay between magnetism and superconductivity.

As suggested from the evolution of entropy, unconventional superconductivity appears to have a magnetic origin. A phenomenological model of magnetically mediated superconductivity derives an attractive pairing potential that increases with the generalized momentum- and frequency-dependent magnetic susceptibility~\cite{monthoux}. Because the magnetic susceptibility diverges at a magnetic quantum-phase transition, the pairing potential will be strongest there, and, indeed, superconductivity in heavy-electron systems typically emerges in proximity to a quantum-phase transition. The nature, however, of the quantum criticality should influence the dominant pairing channel. For example, the susceptibility is singular at the \textbf{Q}-vector defining a spin-density wave that becomes critical at $T=0$, and this favors d-wave pairing~\cite{monthoux}. On the other hand, ferromagnetic criticality favors the p-wave channel. The critical state that emanates from P2 in CeRhIn$_5$ is neither that of an SDW nor of ferromagnetism but is associated with criticality of localized 4f electrons~\cite{tuson08}. Such criticality is consistent with a selective Mott or Kondo-breakdown type of zero-temperature phase transition~\cite{pepin, coleman, si, senthil, paul}. Unlike an SDW quantum-phase transition, fluctuations emerging from this type of unusual quantum-phase transition are not well elaborated but are not expected to be peaked around a particular~\textbf{Q}. In this case, it is not clear what pairing channel might be favored or, from a theoretical perspective, even if superconductivity should appear. Nevertheless, CeRhIn$_5$ appears to provide an example of where this is possible. The simultaneous localized/itinerant roles of the 4f electron in CeRhIn$_5$ seem to be a necessary ingredient in this physics.

\section{Summary}

Specific heat studies of the heavy-fermion compound CeRhIn$_5$ reveal a complex interplay between localized and itinerant degrees of freedom of Ce's 4f electron. The evolution of this interplay with applied pressure is emphasized in the response of CeRhIn$_5$ to applied magnetic field which shows that the 4f electron remains localized up to a critical pressure P2. In spite of its local character, the 4f electron also participates in creating a state of bulk unconventional superconductivity. There are many open problems posed by these discoveries, not the least of which is the need for a theoretical basis for interpreting experimental observations. Understanding the physics of CeRhIn$_5$ should open new perspectives on other complex systems in which electronic correlations entangle states to create emergent functionalities.

\section*{Acknowledgments}
We thank H O. Lee, F. Ronning and V. A. Sidorov for communicating unpublished results and acknowledge numerous helpful discussion with colleagues. TP acknowledges a grant from the Korea Science and
Engineering Foundation (KOSEF) funded by the Korea government R01-2008-
000-10570-0. Work at Los Alamos was performed under the auspices of the US Department of Energy/Office of Science and funded in part by the Los Alamos Laboratory Directed Research and Development program.

\end{document}